\documentclass[preprint,aps,12pt,preprintnumbers,eqsecnum,nofootinbib]{revtex4}
\usepackage{graphicx}
\usepackage{subfigure}

\usepackage{color}
\usepackage{amssymb,amsmath,amsfonts}
\usepackage{epstopdf}

\newcommand{\beq}{\begin{equation}}
\newcommand{\enq}{\end{equation}}

\begin{document}
%
% This version revised for the referee.
%
% Title of paper
\title{\vspace*{0.5in} 
Unitarity and microscopic acausality in a nonlocal theory
\vskip 0.1in}
\author{Christopher D. Carone}\email[]{cdcaro@wm.edu}

\affiliation{High Energy Theory Group, Department of Physics,
College of William and Mary, Williamsburg, VA 23187-8795}
\date{December 14, 2016}
%\date{}

%
%
\begin{abstract}
We consider unitarity and causality in a higher-derivative theory of infinite order, where propagators fall off more quickly in
the ultraviolet due to the presence of a transcendental entire function of the momentum.   Like Lee-Wick theories, these field theories 
might provide new avenues for addressing the hierarchy problem; unlike Lee-Wick theories, tree-level propagators do not have 
additional poles corresponding to unobserved particles with unusual properties. We consider microscopic acausality in these nonlocal theories.
The acausal ordering of production and decay vertices for ordinary resonant particles may provide a phenomenologically distinct 
signature for these models.
\end{abstract}
\maketitle

\section{Introduction} \label{sec:intro}

One path to addressing the hierarchy problem is to consider extensions of the standard model that lessen the degrees of divergence 
of loop integrals.  Historically, supersymmetry has been the most popular approach of this type.  Loop diagrams involving the supersymmetric partners of ordinary particles cancel 
the quadratic divergence of the Higgs boson squared mass that would otherwise be present.  The surviving dependence on any high mass scales in the theory is only logarithmic, so that 
extreme fine tuning is avoided.  In Lee-Wick theories~\cite{lworig}, in particular the Lee-Wick Standard Model~\cite{lwsm}, more convergent loop diagrams are assured by the introduction of 
higher-derivative kinetic terms that yield propagators that fall off more quickly with momentum.   However, a propagator whose inverse is a higher-order polynomial in the  momentum will have additional 
poles.  This fact is reflected in an auxiliary field formulation of Lee-Wick theories in which higher-derivative terms are absent, but additional field are present that correspond to these Lee-Wick partner states~\cite{lwsm}.  Diagrams involving the Lee-Wick partner particles serve to cancel unwanted quadratic divergences, and hence play a role similar to the partner particles in supersymmetric theories.

Among the scenarios with partner particles that address the hierarchy problem, Lee-Wick theories are particularly unusual.  The partner states in Lee-Wick theories have wrong-sign kinetic and mass 
terms, requiring special rules to be applied so that the theory has a chance at a sensible interpretation~\cite{clop}.   However, such states need not appear in all theories with with higher-derivative quadratic 
terms~\cite{nopoles}.  Given the possibility of applications in addressing the hierarchy problem~\cite{Biswas:2014yia}, it is well motivated to consider higher-derivative theories in which the complications of Lee-Wick theories might 
be avoided altogether.

As an example of the type of theory of interest here, consider
\begin{equation}
{\cal L} = - \frac{1}{2} \phi \, \hat{F}(\Box)^{-1}  (\Box+m^2) \, \phi - \frac{\lambda}{4!} \, \phi^4  \,\,\, ,
\label{eq:toy}
\end{equation}
where $\phi$ is a real scalar field, $\Box \equiv \partial_\mu \partial^\mu$ and the momentum-space propagator is given at tree-level by
\begin{equation}
\tilde{D}_F(p) = \frac{i \, \hat{F}(-p^2)}{p^2 -m^2 + i \epsilon} \,\,\, .
\label{eq:toy2}
\end{equation}
If $\hat{F}(-p^2)$ is an entire function, then there will be no additional poles in Eq.~(\ref{eq:toy2}), aside from the one at $p^2=m^2$. 
If $\hat{F}$ is a transcendental function (rather than a polynomial, which is also entire), then we can find forms that drop off at large momentum. In particular, we will 
focus on the simple choice
\begin{equation}
\hat{F}(\Box) = \exp(-\eta\stackrel{}{\Box}^n) \,\, ,
\label{eq:toy3}
\end{equation}
where $\eta>0$ is a coupling constant, and $n$ is a positive, even integer.  (We restrict ourselves to even $n$ so that $\hat{F}$ provides a
convergence factor in either Euclidean or Minkowski space.)  This theory is nonlocal. The consequences of nonlocal modifications
of the quadratic terms in the Lagrangian were discussed as early as the 1950's~\cite{Pais:1950za}, but have periodically met a resurgence of 
interest~\cite{Barnaby:2007yb,Tomboulis:1997gg,Modesto:2011kw,Biswas:2011ar,Biswas:2013cha,Modesto:2014xta,Modesto:2014lga,Biswas:2014yia,Modesto:2015foa,Addazi:2015ppa,Tomboulis:2015gfa}.
Motivated by the infinite-derivative Lagrangians obtained in string field theory~\cite{sft} and $p$-adic string theory~\cite{padic}, nonlocal theories of the general 
type of interest here were studied as possible models of inflation~\cite{Barnaby:2007yb}.   More recently, the possibility that such nonlocal quadratic terms could provide an avenue for quantizing gravity has also been discussed~\cite{Tomboulis:1997gg,Modesto:2011kw,Biswas:2011ar,Biswas:2013cha,Modesto:2014xta,Modesto:2014lga}.  Of particular motivation here is the work of Ref.~\cite{Biswas:2014yia} which applies nonlocal modifications of the quadratic terms to parts of the standard model itself and considers some aspects of the phenomenology.   A more extensive list of background references on nonlocal field theories and their applications can be found in that work.

Ref.~\cite{Biswas:2014yia}, like most phenomenological studies of proposed modifications to the standard model, ultimately focuses on scattering 
processes, which reflect the overlap of asymptotic states defined in the far past and far future.   In the context of Lee-Wick theories, it was pointed out 
by Grinstein, O'Connel and Wise (GOW) that the distinctive acausal features of the theory could be studied by considering the time-dependence of 
the scattering processes via a wave-packet analysis conducted in the semi-classical limit~\cite{Grinstein:2008bg}, as we discuss in more detail later.   
The trajectories of wave packets can be used to define the apparent production and decay points of an exchanged resonance, and the dependence 
of the amplitude on the ordering of these events evaluated.  GOW worked with a theory of real scalar fields with O($N$) symmetry, where the unitarity 
of the theory could be demonstrated to all orders in perturbation theory in the large $N$ limit.   As argued by 
Coleman~\cite{acoleman}, the existence of a unitary $S$-matrix implies that observable acausality does not lead to logical paradoxes in scattering 
experiments, since there is a unitary evolution of initial states to final states.  The question that we wish to study in the present work is how the 
approach and conclusions of GOW are modified if one instead assumes a theory with an infinite-derivative modification of the quadratic terms, one 
that does not introduce additional poles with wrong-sign residues in the propagator at tree-level.

We proceed largely by analogy, first addressing the issue of unitarity in a specific nonlocal extension of the O($N$) model studied by GOW.  Unitarity in nonlocal theories has been discussed in a more general context in Refs.~\cite{Tomboulis:1997gg,Tomboulis:2015gfa,Alebastrov:1973vw} and in a different context in Ref.~\cite{Addazi:2015ppa}.   What we gain by working in the large $N$ limit of the O($N$) model is that unitarity can be studied in explicit detail, to all orders in perturbation theory, via a one-loop calculation. In addition, the intermediate steps and final conclusions can be readily compared to those of 
Ref.~\cite{Grinstein:2008bg}.   The reader who is familiar with the phenomenological work of Ref.~\cite{Biswas:2014yia} will recall that the authors define their nonlocal theory via Euclidean correlation functions that are analytically continued in their external momentum to Minkowski space.  If one were to attempt to quantize the theory in Minkowski space directly, one would find that unitary is violated.  The calculation that we present in Sec.~\ref{sec:unitarity} will make clear why this is the case. We then turn to the issue of causality in Sec.~\ref{sec:causality}.  It is generally expected that the nonlocal theories having the form 
shown in Eq.~(\ref{eq:toy}) have field commutators that do not vanish at space-like separation~\cite{Tomboulis:2015gfa}.  We show that this is indeed the case in the specific O($N$) model defined in Sec.~\ref{sec:unitarity} by an explicit calculation.  What our consideration of unitarity and causality demonstrate up to this point is that the theory of interest may show signs of acausality in scattering experiments without logical inconsistency, in the sense discussed by Coleman.  To address this further, we turn to the scattering of wave packets in the latter half of Sec.~\ref{sec:causality}, and show that there is a non-vanishing amplitude for acausal orderings of production and decay vertices for exchanged resonances.   Unlike the Lee-Wick case, where the resonance is a Lee-Wick partner with wrong-sign kinetic and mass terms, the resonances in this case are ordinary particles.  In Sec.~\ref{sec:conc} we summarize our conclusions.  

The explicit calculations that we present in this work, as well as the discussion of the nonlocal O($N$) model and the detailed application of the approach of Ref.~\cite{Grinstein:2008bg} to similar theories, have not appeared in the 
literature previously.   These may serve as a useful complement to more formal treatments that anticipate the qualitative features of some of our results.  Moreover, the explicit examples and calculations that we present may resonate
with a wider audience of model-builders who are interested in phenomenological applications relevant to TeV-scale physics, an exploration that has been quite limited thus far~\cite{Biswas:2014yia}.

\section{Unitarity} \label{sec:unitarity}

\subsection{Preliminaries}
In the absence of higher-derivative modifications, we work with a theory of $N$ real scalar fields with the Lagrangian density
\begin{equation}
{\cal L} = \frac{1}{2} \partial_\mu \phi^a \partial^\mu\phi^a -\frac{1}{2} m^2 \,\phi^a\phi^a -\frac{1}{8} \lambda_0 (\phi^a\phi^a)^2 \,\,\, .
\label{eq:on1}
\end{equation}
This theory has an O($N$) global symmetry, with the index $a$ running from $1$ to $N$.  The theory has a sensible $N \rightarrow \infty$ limit, {\em i.e.}, there are no Feynman 
diagrams that grow as positive powers of $N$, if the coupling $\lambda_0$  scales as $1/N$. (For a pedagogical discussion, see Ref.~\cite{aspects}.)  It is convenient to redefine 
the coupling $\lambda_0 \equiv  \lambda/N$, so that the $N$ dependence of a given amplitude is explicit.  Following Ref.~\cite{Grinstein:2008bg}, the theory in Eq.~(\ref{eq:on1}) is 
equivalent to
\begin{equation}
{\cal L} = \frac{1}{2} \partial_\mu \phi^a \partial^\mu\phi^a -\frac{1}{2} m^2 \,\phi^a\phi^a + \frac{N}{2 \lambda} \sigma^2 -\frac{1}{2} \sigma \phi^a \phi^a \,\,\, ,
\label{eq:on2}
\end{equation}
where $\sigma$ is an auxiliary field; this can be verified by substitution of the auxiliary field's equation of motion into Eq.~(\ref{eq:on2}).  The advantage of working 
with the auxiliary field formulation is that it makes counting of powers of $N$ transparent, since each $\sigma$ propagator scales as $1/N$.  For example, the self-energy function 
for the $\sigma$ field, $\Sigma_0 (p^2)$, receives it's leading order contribution from a $\phi^a$ loop, and scales as $N$.  Following the sign conventions of
Ref.~\cite{Grinstein:2008bg},  the full $\sigma$ propagator is given by
\begin{equation}
\tilde{D}(p^2) = \frac{i}{1/\lambda_0} +  \frac{i}{1/\lambda_0} \left(i \Sigma_0 \right) \frac{i}{1/\lambda_0}
+ \frac{i}{1/\lambda_0} \left( i \Sigma_0 \right)  \frac{i}{1/\lambda_0} \left(i \Sigma_0 \right)  \frac{i}{1/\lambda_0} + \cdots \,\,\, ,
\end{equation}
which can be re-summed to 
\begin{equation}
\tilde{D}(p^2) =  \frac{\lambda}{N} \frac{i}{1+\lambda \, \Sigma(p^2)} \,\,\, ,
\label{eq:dprop}
\end{equation}
where $\Sigma_0(p^2) \equiv N \,\Sigma(p^2)$, so that the $N$-scaling of Eq.~(\ref{eq:dprop}) is explicit.  All corrections to $\Sigma(p^2)$ that 
are higher than one-loop are suppressed by additional factors of $1/N$, by virtue of the additional $\sigma$ propagators.    Hence, if one is 
interested in only the leading-order behavior of $\Sigma(p^2)$, one only needs to compute a one-loop diagram\footnote{At leading
order there is also a one-loop $\sigma$ tadpole diagram, but it can be eliminated by a shift in the auxiliary field and a redefinition of the $\phi^a$ mass~\cite{aspects}.}.

At leading order in $1/N$, two-into-two scattering in the auxiliary field formulation corresponds to the $s$-, $t$- and $u$- channel exchanges of 
the auxiliary field, with the dressed propagator given by Eq.~(\ref{eq:dprop}).  All other loop corrections to the scattering amplitude involve additional 
$\sigma$ propagators and are sub-leading in the $1/N$ expansion.  It follows that the scattering amplitude is given by
\begin{equation}
{\cal M}(ab\rightarrow cd) = -\frac{\lambda}{N} \left[ \frac{\delta_{ab} \delta_{cd}}{1 + \lambda \Sigma(s)} 
+ \frac{\delta_{ac} \delta_{bd}}{1 + \lambda \Sigma(t)}  
+\frac{ \delta_{ad} \delta_{bc} }{1 + \lambda \Sigma(u)} \right]
\,\,\, ,
\label{eq:scamp1}
\end{equation}
where $s$, $t$ and $u$ are the usual Mandelstam invariants.  As reviewed in Ref.~\cite{Grinstein:2008bg}, Eq.~(\ref{eq:scamp1}) can be used to demonstrate the unitarity of
the theory at leading order in $1/N$ and at all orders in perturbation theory.

Our present interest is how this calculation is altered when there is a nonlocal modification to Eq.~(\ref{eq:on1}), of either the form
\begin{equation}
{\cal L} =
- \frac{1}{2} \phi^a \, \hat{F}(\Box)^{-1}  (\Box+m^2) \, \phi^a -\frac{1}{8} \lambda_0 (\phi^a\phi^a)^2 \,\,\, ,
\label{eq:nloc2}
\end{equation}
or
\begin{equation}
{\cal L} = - \frac{1}{2} \phi^a \,  (\Box+m^2) \, \phi^a  
-\frac{1}{8} \lambda_0 \left[(\hat{F}^{1/2}\phi^a)( \hat{F}^{1/2}\phi^a)\right]^2 \,\,\, .
\label{eq:nloc1}
\end{equation}
Here $\hat{F}$ is the differential operator defined in Eq.~(\ref{eq:toy3}), with $\eta>0$ and $n$ and even positive integer, and 
$\hat{F}^{1/2} \equiv \exp(-\frac{1}{2} \eta\!\stackrel{}{\Box}^n)$. We choose $n$ even
so that the factors of $\hat{F}$ lead to improved convergence of loop integrals in momentum space, regardless of whether we assume
a Euclidean or Minkowski metric.  We compare each possibility in the following subsection, for the simplest choice of $n=2$, which we assume henceforth.  
Eqs.~(\ref{eq:nloc2}) and (\ref{eq:nloc1}) are related by a nonlocal field redefinition and give the same results for scattering
amplitudes. Working with Eq.~(\ref{eq:nloc1}), the factors of $\hat{F}^{1/2}$ acting on internal lines reproduce the momentum dependence of the propagator that 
one obtains from Eq.~(\ref{eq:nloc2}); the factors of $\hat{F}^{1/2}$  acting on external lines each give a factor of $\exp(-\eta \, m^4/2)$, matching the wave function renormalization factors in the 
scattering amplitudes obtained from Eq.~(\ref{eq:nloc2}).   For definiteness, we will examine the case 
where $a=b \neq c=d$ so that only the $s$-channel amplitudes is relevant.  Then the scattering amplitude takes the form
\begin{equation}
{\cal M} = - \frac{\lambda}{N}  \frac{e^{-2 \eta \, m^4}}{1 + \lambda\, \Sigma(s)} \, \delta_{ab} \delta_{cd} \,\,\, ,
\label{eq:ourm}
\end{equation}
where the constant exponential factor is due to the higher-derivative operator acting on the external lines, and where $\Sigma(s)$ now includes the 
effects of $\hat{F}$ on the $\phi^a$ propagator. 

\subsection{Minkowski Space}

We show in this section that the theory defined in Minkowski space by Eq.~(\ref{eq:nloc1}) violates unitarity.   The self-energy function $\Sigma(p^2)$ is
given by
\begin{equation}
\Sigma(p^2) = -\frac{i}{2} \int\frac{d^4k}{(2 \pi)^4} \frac{\exp\{-\eta \, (k+p/2)^4\}\exp\{-\eta \,(k-p/2)^4\}}{[(k-p/2)^2 -m^2 + i \epsilon] [(k+p/2)^2-m^2 + i \epsilon]} \,\,\, .
\label{eq:sigmink}
\end{equation} 
Unitarity implies the operator relation $i (T^\dagger-T) = T^\dagger T$, where the $T$-matrix is related to the $S$-matrix by $S=1+i T$.  One can derive a condition on
scattering amplitudes by taking matrix elements of both sides of this expression and including an appropriate insertion of a complete set of intermediate states.  Working at leading 
order in the $1/N$ expansion, this procedure was carried out in the O(N) model in Ref.~\cite{Grinstein:2008bg}, and the derivation is not altered by the presence of the
additional momentum space suppression factors in the numerator of Eq.~(\ref{eq:sigmink}).   One finds~\cite{Grinstein:2008bg}
\begin{eqnarray}
 2 \mbox{ Im}&\!\!{\cal M}& \!\!(k_1,a; k_2,b \rightarrow k_1',c;k_2',d) = 
 \sum_{e,f} I_{e,f} \int \frac{d^3 q_1}{(2 \pi)^3} \frac{d^3 q_2}{(2 \pi)^3} \frac{1}{2 E_1}\frac{1}{2 E_2} (2 \pi)^4 \delta^{(4)}(q_1 + q_2 -p) \nonumber \\ &&
 {\cal M}(k_1,a; k_2,b \rightarrow q_1,e ; q_2, f) \, {\cal M}^*(k_1', c; k_2' , d \rightarrow q_1, e; q_2, f) \,\,\, ,
\label{eq:opthm}
\end{eqnarray}
where the identical particle factor $I_{e,f}=1/2$ if $e=f$ and $1$ otherwise.  The left-hand-side of this expression follows immediately from Eq.~(\ref{eq:ourm}):
\begin{equation}
LHS = \frac{\lambda^2}{N} \left[2 \,e^{-2 \, \eta \, m^4} \mbox{ Im }\Sigma(s) \right]  \frac{1}{|1+\lambda\, \Sigma(s)|^2}  \,\delta_{ab} \delta_{cd}\,\,\, .
\label{eq:lhs}
\end{equation}
The right-hand-side of Eq.~(\ref{eq:opthm}) includes only two-particle intermediate states, which provide the leading contribution in the large $N$ limit.  After
substitution of Eq.~(\ref{eq:ourm}), the necessary integral evaluation is identical to that of the two-body Lorentz-invariant phase space factor.   The result is
\begin{equation}
RHS = \frac{\lambda^2}{N} \left[ \frac{1}{16 \pi} e^{-4 \, \eta \, m^4} \sqrt{1-\frac{4 m^2}{s}} \,\right] \frac{1}{|1+\lambda\, \Sigma(s)|^2} \,\delta_{ab} \delta_{cd}  \,\,\, .
\label{eq:rhs}
\end{equation}
When $\eta=0$, the quantities in brackets in Eqs.~(\ref{eq:lhs}) and (\ref{eq:rhs}) coincide, as can be seen either from an elementary one-loop calculation~\cite{Grinstein:2008bg}, or by examining the $\eta \rightarrow 0$ limit of the numerical calculation that we are about to describe.  When
$\eta \neq 0$, these quantities differ.  After exploring the source of the discrepancy, we show how it is avoided by defining the theory as an
analytic continuation of correlation functions defined in Euclidean space.

\begin{figure}[t]
  \begin{center}
    \includegraphics[width=0.5\textwidth]{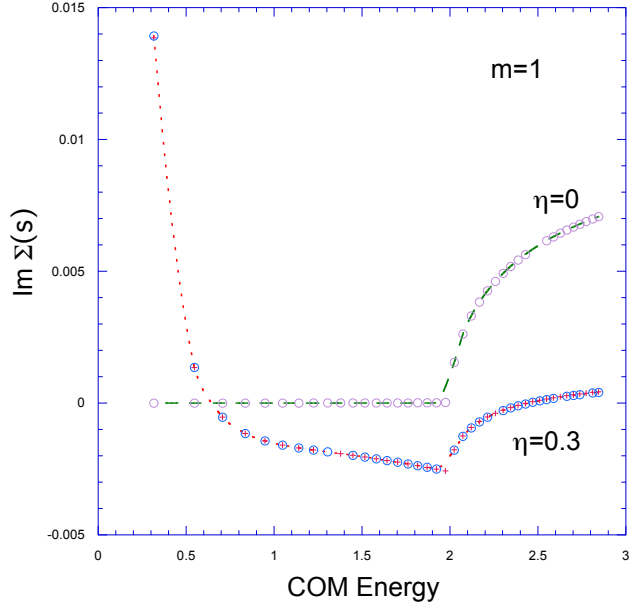}
    \caption{ Imaginary part of the self-energy functions $\Sigma(s)$ as a function of $\sqrt{s}$. The open circular points indicate the results of the direct
    numerical evaluation of Eq.~(\ref{eq:sigmink}), for the cases where $\eta=0$ and $\eta=0.3$.  The long dashed line gives
    the $\eta=0$ expectation, proportional to $(1-4 \, m^2 /s)^{1/2}$ for $s >4 m^2$.  The solid points are 
    the result of Eqs.~(\ref{eq:res1}), after a numerical evaluation of the second term, as discussed in the text.}
        \label{fig:ufig}
  \end{center}
\end{figure}

It is easiest to see that Eqs. (\ref{eq:lhs}) and (\ref{eq:rhs}) do not agree when $\eta \neq 0$ by showing that $\mbox{Im } \Sigma (p^2)$
no longer has a functional form proportional to $\sqrt{1- 4\, m^2/ s}$.  To confirm this claim most quickly, we simply evaluate the imaginary part of Eq.~(\ref{eq:sigmink}) numerically, working in the center-of-mass frame, where $\vec{p}=0$;  we perform the $k^0$ integral exactly along the 
real axis with $\epsilon$ finite and evaluate the limit as $\epsilon \rightarrow 0$.  Note that the imaginary part of the loop integral is finite, even when
$\eta$ is vanishing. It is convenient to write Eq.~(\ref{eq:sigmink}) in the following form:
\begin{equation}
\mbox{Im } \Sigma =-\frac{1}{4 \pi^3} \int_0^\infty dk \int_0^\infty dk^0 \left\{ k^2 g(k^0,k)
\frac{f_+(k^0,k) f_-(k^0,k) - \epsilon^2}{\left[f_+(k^0,k)^2+\epsilon^2\right] \left[f_-(k^0,k)^2+\epsilon^2\right]} \right\}  \,\,\, ,
\end{equation}
where $k \equiv |\vec{k}|$, 
\begin{equation}
f_\pm(k^0,k) \equiv (k^0 \pm p^0/2)^2-k^2-m^2 \,\,\, ,
\end{equation}
and
\begin{equation}
g(k^0,k) \equiv \exp\{-\eta [(k^0 + p^0/2)^2-k^2]^2-\eta [(k^0 - p^0/2)^2-k^2]^2 \} \,\,\, .
\end{equation}
The integration can be performed using symbolic mathematics code (we used MAPLE~\cite{maple}), provided care is taken in dealing with the points on the real
$k^0$ axis that would be singularities in the $\epsilon \rightarrow 0$ limit.   For $\epsilon$ small but non-zero, the growth of the integrand around these points
are taken into account by singularity handling routines in MAPLE that are invoked automatically by breaking up the region of $k^0$ integration into intervals that are terminated at these points.  We then have no difficulty obtaining numerically convergent results.   In Fig.~\ref{fig:ufig}, we show the result for $\mbox{Im } \Sigma$ as a 
function of the center-of-mass energy, working in units where $m=1$, for $\eta=0$ and an example where $\eta \neq 0$.  The line with long dashes shows the 
expectation for $\mbox{Im } \Sigma$ following from the analytic result of the one-loop calculation in the $\eta=0$ case,
\begin{equation}
\mbox{Im } \Sigma (s) = \frac{1}{32 \pi} \sqrt{1-\frac{4 m^2}{s}} \theta(s -4 m^2)  \,\,\, ,
\end{equation}
where $\theta$ is the Heaviside step function.   This agrees with the numerical result for $\eta=0$, given by the open circular points in Fig.~\ref{fig:ufig}.
However, the results are not proportional to the same functional form in $s$ for the case where $\eta \neq 0$.  One would not suspect that the disagreement is the 
result of a numerical artifact, since the extra exponential factor in the integrand in the case where $\eta \neq 0$ is smooth and serves primarily to 
truncate the domain of integration.  

To further verify this result, let us now do the calculation in a different way. Imagine we evaluate the $k^0$ integral in $\Sigma$ by closing a semi-circular contour in the lower-half complex plane.  In ordinary, local 
quantum field theories, the integral along the semi-circular contour would vanish as the radius of the contour is taken to infinity.  In the present theory, this is not the case; the numerator of the loop integral becomes 
$\exp[-2 \, \eta\, (k^0)^4]$ far from the origin, which blows up in directions where $\mbox{Re } (k^0)^4 <0$.   Hence, let us decompose
\begin{equation}
\Sigma =  \Sigma_p -  I_C \,\,\,\, \mbox{ and } \,\,\,\, 2 \, \mbox{Im} \, \Sigma = 2\, \mbox{Im} \, \Sigma_p - 2\, \mbox{Im} \, I_C \,\,\, ,
\label{eq:decomp}
\end{equation}
where $\Sigma_p$ is $-2 \pi i$ times the residues of the poles contained within the contour and $I_C$ is the clockwise integral along the semi-circular portion.  Since
\begin{equation}
\Sigma_p(p^0) = \frac{1}{2} \int \frac{d^3 k}{(2 \pi)^3} \frac{1}{2 E_{\vec{k}}\,p^0} \left[ \frac{N(k^0 = E_{\vec{k}} - p^0/2)}{2 E_{\vec{k}} - p^0}
-\frac{N(k^0 = E_{\vec{k}} + p^0/2)}{2 E_{\vec{k}} + p^0} \right] \,\,\, ,
\label{eq:ppart}
\end{equation}
where $N$ represents the numerator of the integrand of Eq.~(\ref{eq:sigmink}) and $E^2_{\vec{k}}\equiv |\vec{k}|^2+m^2$, it is clear
for $0<p^0<2 m$ on the real axis that $\Sigma_p(p^0) = \Sigma_p(p^{0*})^*$.  Since $\Sigma_p$ is an analytic function of $p^0$ off the real axis, this can be analytically continued, from which it follows that
\begin{equation}
\mbox{Disc} \, \Sigma_p = 2 \, i \, \mbox{Im} \,\Sigma_p (p^0+i \epsilon) \,\,\, ,
\label{eq:discim}
\end{equation}
where the discontinuity is about a cut on the real $p^0$ axis,
\begin{equation}
\mbox{Disc}\, \Sigma_p = \lim_{\epsilon \rightarrow 0} \left[ \Sigma_p(p^0+i \epsilon) -\Sigma_p(p^0-i\epsilon) \right] \,\,\, .
\end{equation}
We may express
\begin{eqnarray}
\Sigma_p(p^0+i \epsilon) &=&\frac{1}{8 \pi^2} \int d E_{\vec{k}} \frac{ |\vec{k}|}{p^0 + i \epsilon} \nonumber \\
&&  \times
\left[ \frac{N(k^0 = E_{\vec{k}} - p^0/2 -i \epsilon/2)}{2 E_{\vec{k}} - p^0-i \epsilon}-\frac{N(k^0 = E_{\vec{k}} + p^0/2+i \epsilon/2)}{2 E_{\vec{k}} + p^0+i \epsilon} \right] \,\,\, .
\end{eqnarray}
There can be poles in the $E_{\vec{k}}$ integration that blow up at most as $1/\epsilon$;  hence, we only need expand what remains  
to order $\epsilon$.  Defining $N' \equiv \partial N / \partial k^0$, we find
\begin{equation}
\begin{split}
\mbox{Disc}\, \Sigma_p & = - \frac{1}{8\pi^2} \lim_{\epsilon \rightarrow 0}  \int d E_{\vec{k}} \frac{|\vec{k}|}{p^0}  \\
& \left\{ N(k^0 = k^0_-) \left[\frac{1}{p^0-2 E_{\vec{k}} +i \epsilon}- \frac{1}{p^0-2 E_{\vec{k}} - i \epsilon}\right] \right. \\
&+ N(k^0 = k^0_+) \left[\frac{1}{p^0+2 E_{\vec{k}} +i \epsilon}- \frac{1}{p^0+2 E_{\vec{k}} - i \epsilon}\right]  \\
& +\left[\frac{-i}{2} N'(k^0 = k^0_-)-\frac{i}{p^0}N(k^0 = k^0_-)\right]
 \left[\frac{\epsilon}{p^0-2 E_{\vec{k}} +i \epsilon}+ \frac{\epsilon}{p^0-2 E_{\vec{k}} - i \epsilon}\right]  \\
& \left.+\left[\frac{i}{2} N'(k^0 = k^0_+)-\frac{i}{p^0}N(k^0 = k^0_+)\right] \left[\frac{\epsilon}{p^0+2 E_{\vec{k}} +i \epsilon}+ \frac{\epsilon}{p^0+2 E_{\vec{k}} - i \epsilon}\right] \right\}
\end{split}
\end{equation}
where $k^0_\pm = E_{\vec{k}} \pm p^0/2$.  We can now take the $\epsilon \rightarrow 0$ limits of the quantities in square brackets, using
\begin{equation}
\lim_{\epsilon \rightarrow 0} \left[\frac{1}{y+i \epsilon} - \frac{1}{y-i \epsilon}  \right] = -2 \pi i\, \delta(y)
\,\,\, \mbox{ and } \,\,\,
\lim_{\epsilon \rightarrow 0} \left[\frac{\epsilon}{y+i \epsilon} + \frac{\epsilon}{y-i \epsilon} \right] = 2 \pi y \, \delta(y) \,\,\, .
\end{equation}
We see that the third and fourth terms in the curly braces are proportional to $(p^0 \pm 2 E_{\vec{k}}) \delta(p^0 \pm 2 E_{\vec{k}})$,
so that they vanish after integration.  Since $p^0>0$, the surviving term is given by
\begin{equation}
\mbox{Disc}\, \Sigma_p  = i \frac{1}{4 \pi} \int d E_{\vec{k}}\, \frac{|\vec{k}|}{p^0} \, \delta(p^0 - 2 E_{\vec{k}}) \, N(k^0=E_{\vec{k}}-p^0/2)
\end{equation}
It is straightforward to confirm that the same result is obtained by making the conventional Cutkosky replacements in
the original integral for $\Sigma_p$
\begin{equation}
\frac{1}{(k \pm p/2)^2-m^2 + i \epsilon} \rightarrow -2 \pi i \,\delta([k \pm p/2]^2-m^2)
\end{equation}
so that
\begin{equation}
i \, \mbox{Disc} \, \Sigma_p = \frac{1}{2} \int \frac{d^4 k }{(2 \pi)^4} N(k) (-2 \pi i)^2 \delta([k + p/2]^2-m^2)) \delta([k - p/2]^2-m^2) \,\,\, .
\end{equation}
Changing variables, introducing an additional integral, and writing out the numerator factor $N$, this is equivalent to
\begin{equation}
\begin{split}
i \, \mbox{Disc} \, \Sigma_p &=  \\
   -\frac{1}{2} &\int \frac{d^4 q_1}{(2 \pi)^4} \frac{d^4 q_2}{(2 \pi)^4} \, e^{(-\eta \, q_1^4-\eta \,q_2^4)}  \,
 (2 \pi) \delta(q_1^2-m^2) \, (2 \pi) \delta(q_2^2-m^2)  \\
&  (2 \pi)^4 \delta^{(4)}(q_1+q_2-p) \,\,\, , 
\end{split}
\end{equation}
which integrates to
\begin{equation}
i \, \mbox{Disc} \, \Sigma_p = -\frac{1}{16 \pi} e^{-2 \,\eta\, m^4} \sqrt{1-\frac{4 m^2}{s}}  \,\,\, .
\end{equation}
From Eqs.~(\ref{eq:decomp}) and (\ref{eq:discim}), it follows that we can write the quantity in square brackets from
the left-hand-side of our unitarity relation, Eq.~(\ref{eq:lhs}), as
\begin{equation}
\left[ 2 \,e^{-2 \, \eta \, m^4} \mbox{ Im }\Sigma(s) \right]= \frac{1}{16 \pi} e^{-4 \,\eta\, m^4} \sqrt{1-\frac{4 m^2}{s}} - 2\, e^{-2 \,\eta\, m^4}
 \, \mbox{Im} \, I_C  \,\,\,.
 \label{eq:res1}
\end{equation} 
The first term agrees with the desired form of the quantity in square brackets in Eq.~(\ref{eq:rhs});  it follows that the violation of unitarity stems entirely 
from the non-vanishing of the integral $I_C$ along the semi-circular contour.  

We can verify that Eq.~(\ref{eq:res1}) is correct by evaluating the imaginary part of $I_C$ and comparing the result for
$\mbox{Im} \,\Sigma$ with what we obtained previously in Fig.~\ref{fig:ufig}.  Notice that if we were to push all the poles on the real $k^0$ axis to
the upper half-plane, then $I_C$ would be given by the negative of the integral along the real axis.  Hence, we may identify
\begin{equation}
I_C = \frac{i}{2} \int \frac{d^4 k}{(2 \pi)^4} \frac{N(k)}{\left[(k^0-p^0/2 - i \epsilon)^2 - E_{\vec{k}}^2\right]\left[(k^0+p^0/2-i \epsilon)^2- E_{\vec{k}}^2 \right]}  \,\,\, .
\label{eq:icf2}
\end{equation}
The point is that Eq.~(\ref{eq:icf2}) can be evaluated numerically in exactly the same way as the integral in Eq.~(\ref{eq:sigmink}) that we described earlier.  The result for 
$\mbox{Im}\,\Sigma$ computed from Eq.~(\ref{eq:res1}) using the numerical evaluation of Eq.~(\ref{eq:icf2}) is indicated by the solid points shown in Fig.~\ref{fig:ufig}: they are in complete agreement 
with our previous direct evaluation of $\mbox{Im}\,\Sigma$ in the case where $\eta \neq 0$.

To understand this result, it is useful to consider how the calculation might have proceeded had we started by evaluating the discontinuity of 
Eq.~(\ref{eq:sigmink}) directly using Cutkosky's formula~\cite{Cutkosky:1960sp}.   It is straightforward to check that the discontinuity computed in this way 
would reconcile Eq.~(\ref{eq:lhs}) and Eq.~(\ref{eq:rhs}) only if $\mbox{Disc}\, \Sigma = 2 \, i \,\mbox{Im} \,\Sigma$.   However, this relation is not justified 
in the present case since $\Sigma$ cannot be shown to satisfy the Schwartz reflection principle $\Sigma(p^0) = \Sigma(p^{0*})^*$ when $\eta \neq 0$.   
The reflection principle requires that there be a segment along the real $p^0$ axis over which $\Sigma$ is purely real;  in the case where $\eta \neq 0$ 
it is not possible to prove that such a region exists and our numerical results shown in Fig.~\ref{fig:ufig} strongly suggest that the opposite is true.   In 
the Appendix, we show in more detail how the violation of the Scwartz reflection principle can be directly related to the non-vanishing of contour integrals, 
like $I_C$, at large radius in the complex plane.

Another starting point~\cite{Tomboulis:1997gg} for attempts to demonstrate unitarity is the Largest Time Equation (LTE)~\cite{Veltman:1963th}.  We simply note here that this approach 
cannot be consistently applied to the present problem.   As discussed by Anselmi~\cite{Anselmi:2016fid}, derivation of the LTE requires two 
assumptions: (1) the vertices of the theory are localized time and (2) the propagator in position space is of the form 
$\theta(x^0) g_+(x) + \theta(-x^0) g_-(x)$, where $\theta$ is the step function.  If nonlocality appears in the vertices of the theory, then assumption (1) is violated.  If a field redefinition is used to move the 
nonlocality to the propagators, then assumption (2) is violated due to the appearance of additional terms in the propagator that are proportional to derivatives of $\delta(x^0)$.  (The explicit form of the 
propagator can be found in Ref.~\cite{Tomboulis:2015gfa}.)  The subsequent derivation of the LTE described in Ref.~\cite{Anselmi:2016fid} fails.  Hence, we say nothing further 
about this approach.

\subsection{Euclidean Space}

We have discussed in the previous subsection how unitarity is violated if we attempt to formulate the theory of interest directly in Minkowski space.  If correlation functions are
defined in Euclidean space and analytically continued in the external momenta to Minkowski space, unitarity is preserved. This might be expected since the theory quantized via
a Euclidean functional integral automatically satisfies reflection positivity.   The way that the calculation of the previous section is modified is as follows:  The Euclidean version of
$\Sigma$ corresponds to Eq.~(\ref{eq:sigmink}) with the $k^0$ integration taken along the imaginary axis, and with Euclidean external momentum $p^0=i p^0_E$.  In other words, 
the starting point is the path that one would obtain with a Wick rotation if it were justified in a Minkowski-space formulation of the theory.   Now, close the contour with a semi-circle 
in the right half plane, so that 
\begin{equation}
2\, \mbox{Im} \, \Sigma = 2\, \mbox{Im}\, \Sigma_p - 2\, \mbox{Im}\, I_C'  \,\,\, ,
\label{eq:decomp2}
\end{equation}
where $I_C'$ is the integral over the semi-circular path, and $\Sigma_p$ again is determined by the residues of the poles contained within the closed contour.   While $\mbox{Im}\, I_C$  in our previous calculation was non-vanishing, we now show that $\mbox{Im}\, I_C'=0$.   Let us write $I_C' = -i \int dk^0 X(k^0)$, where $X(k^0)$ is given by Eq.~(\ref{eq:sigmink}) with $\epsilon$ set to zero and $k^0$ placed on the desired semi-circle, $k^0 = \rho \exp( i \theta)$ for $-\pi/2 \leq \theta \leq \pi/2$.  In the 
center-of-mass frame where $\vec{p}=0$, it is straightforward to check that $X(k^0)$ is also a function of ${p^0}^2={-{p^0_E}^2}$, which is real; it follows immediately that 
$X(k^0)^* = X(k^{0*})$. Since $dk^0 = i k^0 d\theta$,
\begin{equation}
2\, i \, \mbox{Im}\, I_C'=I_C'  - I_C'^*= \lim_{\rho \rightarrow \infty} \int_{-\pi/2}^{\pi/2} d\theta \left[ k^0 X(k^{0}) - k^{0*} X(k^{0*}) \right] \,\,\, ,
\end{equation}
which vanishes; this can be seen by changing variables $\theta \rightarrow -\theta$, and noting that $k^0(-\theta) =k^{0*}$, indicating that the $d\theta$ integral is equal to its negative. The surviving term in Eq.~(\ref{eq:decomp2}) is the same function of $p^0$ that reconciled the left- and right-hand-sides of our unitarity relation in the previous section.
By Lorentz invariance, the result holds in any other reference frame in which the scattering process is measured. Hence, we have verified that the large $N$ scattering amplitudes of interest in the present context are unitary provided that the theory is defined as in Ref.~\cite{Biswas:2014yia}, via the analytic continuation to Minkowski-space external momentum of correlation functions defined in a Euclidean field theory.  We will assume that correlation functions are computed in this way in the discussion that follows.

\section{Causality} \label{sec:causality}

Nonlocal theories of the type studied here were known long ago to violate causality~\cite{Pais:1950za}.  In general, the commutator of fields at space-like 
separation is expected to be non-vanishing for theories where $\hat{F}$ is an entire, transcendental function~\cite{Tomboulis:2015gfa}.  We demonstrate this 
in the case where $m=0$ in the unitary theory discussed in the previous section, a limit in which we can explicitly evaluate the commutator.  We will then turn to 
scalar theories with similar nonlocal modifications and consider how acausality affects the time-dependence of scattering amplitudes, following the general approach of Ref.~\cite{Grinstein:2008bg}.

\subsection{Commutator}
\begin{figure}[t]
  \begin{center}
    \includegraphics[width=0.5\textwidth]{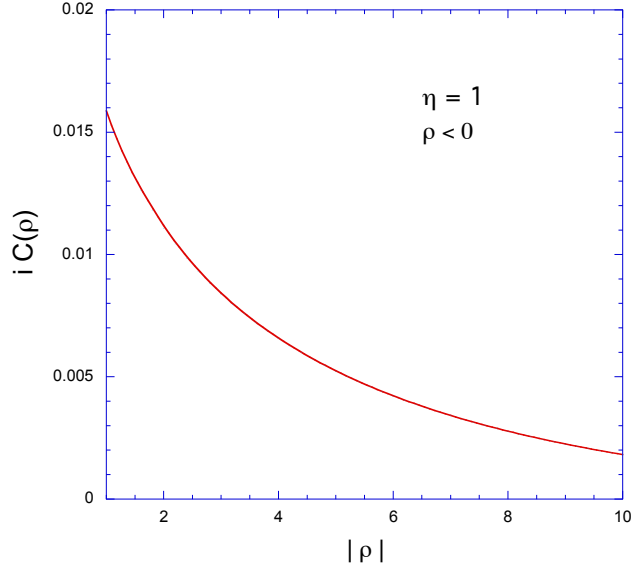}
    \caption{The commutator $C(\rho) = \langle 0 | [\phi(x) ,  \phi(y)] | 0 \rangle$, as a function of
    $\rho = (x^0-y^0)^2-|\vec{x}-\vec{y}|^2$  at space-like separation $\rho<0$, for $m=0$ and $\eta=1$.}
        \label{fig:com}
  \end{center}
\end{figure}
The Feynman-propagator $D_F(x-y)$ is identified with the two-point correlation function $\langle 0 | T \phi(x) \phi(y) | 0 \rangle$.  If we
strictly assume that $x^0>y^0$, then we may write the commutator
\begin{equation}
\langle 0 | [\phi(x) ,  \phi(y)] | 0 \rangle = D_F(x-y) - D_F(x-y)^*  \,\,\,\,\,\,\,\,\,\, (x^0>y^0) \,\,\, .
\label{eq:dfdf}
\end{equation}
Working with the form of the theory in which the higher-derivative operator appears in the quadratic terms for $\phi$, it
follows immediately that\footnote{Since we work here with the lowest-order propagator, our prescription of starting
with a Euclidean correlation function and continuing to Minkowski space in the external momentum simply gives us the
usual momentum-space propagator.  All subsequent Fourier transforms are, of course, in Minkowski space.} 
\begin{equation}
D_F(x-y) = \int \frac{d^4 k}{(2 \pi)^4} \frac{i \, e^{-\eta \, k^4}}{k^2 -m^2 +i \epsilon} e^{-i k \cdot (x-y)}  \,\,\, .
\label{eq:dfdef}
\end{equation}
Because the factor of $e^{-\eta \, k^4}$ blows up in certain directions in the complex $k^0$ plane, as indicated earlier, the usual procedure
of closing the integration contour in the lower half plane is not useful; instead we directly evaluate the $k^0$ integral along the real axis, deviating
by small semi-circles below and above the poles at $k^0 = -E_{\vec{k}}$ and $+E_{\vec{k}}$, respectively.  Hence, we 
may write $D_F(x-y) = I_{2C} + I_{PV}$, where $I_{2C}$ is the result from the semi-circle integrations while $I_{PV}$ is the
remaining principal value integral.   The latter can be reduced to a one-dimensional integral in the case $m=0$, which we can 
numerically evaluate.   We find
\begin{equation}
I_{2C} = \frac{1}{4 \pi^2 |\vec{r}|} \int_0^\infty dk \sin(k\, |\vec{r}|) \cos(k \,r^0) \,\,\, ,
\end{equation}
where we define $r=x-y$ and $k=|\vec{k}|$.  The remaining principal value integral is identical to one considered in the computation of the retarded
propagator for this theory in Ref.~\cite{Pais:1950za}, and is given by
\begin{equation}
I_{PV} = \frac{-i}{\pi^3} \frac{\partial}{\partial \rho} \left[ \mbox{sign}(\rho) \int_0^\infty \frac{dy}{y}  \exp(-\eta y^4/\rho^2)
\left[K_0(y)+\frac{\pi}{2} Y_0(y)\right] \right] \,\,\, .
\label{eq:ipv}
\end{equation}
where $\rho = {r^0}^2-|\vec{r}|^2$. Restricting to the case where $r^0 < |\vec{r}|$, it follows from Eqs.~(\ref{eq:dfdf}) and (\ref{eq:ipv})
that
\begin{equation}
\langle 0 | [\phi(x) ,  \phi(y)] | 0 \rangle = \frac{4 \, i\, \eta}{\pi^3 \rho^3} \int_0^\infty dy \, y^3 \exp(-\eta \,y^4/\rho^2)
\left[K_0(y)+\frac{\pi}{2} Y_0(y)\right]   \,,\,\,\,\,\,\,\, (0< r^0 < |\vec{r}|) \,\,\, ,
\label{eq:comres}
\end{equation}
where $K_0$ and $Y_0$ are Bessel functions.  Eq.~(\ref{eq:comres}) is nonvanishing, as is shown in Fig.~\ref{fig:com}.  We do not consider the case 
$m \neq 0$, since the necessary two-dimensional numerical integration is much more tedious but no more illuminating.

One might wonder how this calculation of the commutator relates to a similar calculation in the formulation of the theory where
the nonlocality appears only in the interaction terms, Eq.~(\ref{eq:nloc1}).   The unstated assumption is that the theory presented
in this form results from a field redefinition starting with the theory in which only the quadratic terms are modified, Eq.~(\ref{eq:nloc2}).  
With nonlocal interactions arising in this way, operators that correspond to observables are built out of the ``smeared" fields, 
$\hat{F}^{1/2} \phi(x)$, and it is the commutator of these objects that is the physically relevant quantity to evaluate at space-like separation.  
This gives precisely the same integral to evaluate as in Eq.~(\ref{eq:dfdef}), with a different origin for the momentum dependence in the 
numerator.

\subsection{Acausal Vertex Ordering}

The question we now wish to address is how acausality manifests itself in the time-dependence of scattering processes.  We allow ourselves 
to stray from the O($N$) model in this subsection to consider theories of a single real scalar field with the same modification of their quadratic 
terms as Eq.~(\ref{eq:nloc2}), but with different interaction terms.  This will allow us to illustrate the effects of interest most clearly; the
generalization to the O($N$) model that we previously considered will be clear by analogy.   We consider two examples, following
the general approach of Ref.~\cite{Grinstein:2008bg}:

{\em Particle Production by a Source.}   Consider a local theory of a real scalar field which includes a coupling to a classical source,
\begin{equation}
{\cal L }_{int} = \phi(x) j(x) \,\,\, ,
\end{equation}
where ${\cal L}_{int}$ is the interaction Lagrangian. We wish to study $\langle \psi_{out} | \Omega \rangle$, the amplitude for the source to create an 
outgoing wave-packet state from the vacuum, where
\begin{equation}
| \psi_{out} \rangle = \int d^4 x' g(x'-y') \, \phi(x') \, | \Omega \rangle  \,\,\, .
\end{equation}
Here we follow the convections of Ref.~\cite{Grinstein:2008bg} where primed coordinates correspond to ``out" states.  The function $g(x')$ is chosen so that the outgoing wave packet is localized about the space-time point $y'$ at some 
time long after the source is turned off, and its four-momentum is localized about $p'$.   By the choice of this function, we can determine the position of the wave-packet at any earlier time when the source is turned on.  For a source localized within a small region about the spacetime origin, we first show that the amplitude vanishes if the wave-packet's trajectory extrapolates back to the origin at a time substantially earlier than $t=0$, as one would expect for a causal process.  We then consider how this conclusion changes given the chosen nonlocal modification  of this theory.  

The amplitude $\langle \psi_{out} | \Omega\rangle$ may be written
\begin{eqnarray}
\langle \psi_{out} | \Omega\rangle &=& \int d^4 x' g^*(x'-y') \langle \Omega | \phi(x') | \Omega \rangle \nonumber \\
&=& i \int d^4 y \, j(y) \int \frac{d^4 p}{(2 \pi)^4} \frac{i}{p^2-m^2+i \epsilon} \, \tilde{g}(p)^* \, e^{-i p \cdot (y'-y)} \nonumber \\
&\equiv& i \int d^4 y \, j(y) \, I(y'-y)  \,\,\,,
\end{eqnarray}
where $\tilde{g}(p)$ is the Fourier transform $\tilde{g}(p)= \int \!d^4 x \, g(x)\, e^{i p\cdot x}$.   As we discussed earlier, all momentum-space
correlation functions are defined via analytic continuation from a Euclidean theory; all subsequent calculations, 
including Fourier transforms, are then performed in Minkowski space. The integral $I(y'-y)$ can be re-expressed using a Schwinger parameter,
\begin{equation}
I(\Delta y) = \frac{1}{\hbar} \int_0^\infty ds \int \frac{d^4 p}{(2 \pi)^4}   \, e^{i s (p^2-m^2 + i \epsilon)/\hbar} \,  
\tilde{g}(p)^* \, e^{-i p\cdot \Delta y/\hbar}  \,\,\, ,
\label{eq:schwing}
\end{equation}
where $\Delta y \equiv y'-y$, and we have temporarily restored the $\hbar$ dependence.  As in Ref.~\cite{Grinstein:2008bg},
if the relevant distance scales (in this case $\Delta y$) are large compared to all characteristic inverse masses and inverse momenta, then we
are justified in using the stationary phase approximation, since this limit is equivalent to taking $\hbar \rightarrow 0$ in Eq.~(\ref{eq:schwing}). 
Evaluating the $d^4 p$ integral in this way (and resuming our convention that $\hbar=1$) yields
\begin{equation}
I(\Delta y) = \frac{i}{16 \pi^2} \int_0^\infty ds\, \frac{1}{s^2} \, \tilde{g}(\frac{\Delta y}{2 s})^* \, e^{-i [ \Delta y^2/(4s) + s m^2]} \,\,\,.
\end{equation}
By evaluating the $ds$ integral in the same way one finds
\begin{equation}
I(\Delta y) = \frac{\sqrt{i}}{4 \sqrt{2} \pi^{3/2}} \frac{m^{1/2}}{(\Delta y^2)^{3/4}} \tilde{g}(m \frac{\Delta y}{\sqrt{\Delta y^2}})^* e^{-i m \sqrt{\Delta y^2}} \,\,\, ,
\end{equation}
leading finally to
\begin{equation}
\langle \psi_{out} | \Omega \rangle = \frac{1}{2} \left(\frac{i}{2\pi}\right)^{3/2} \int d^4 y \, \frac{m^{1/2}}{[(y'-y)^2]^{3/4}} \, j(y) \,
\tilde{g} \!\left( m \frac{y'-y}{\sqrt{(y'-y)^2}} \right)^* e^{-i m \sqrt{(y'-y)^2}}  \,\,\, .
\label{eq:theamp}
\end{equation}
By construction, the function $\tilde{g}$ only has support in the region where 
\begin{equation}
m \frac{(y' - y)} {\sqrt{(y'-y)^2}} \approx p'  \,\,\, ,
\end{equation}
which limits the possible values of $y$ that contribute to the integral.   Let us assume a $\tilde{g}$ in which $\vec{y} \approx 0$ for some $y^0  \ll 0$.   If $j(y)$
is strongly localized about the spacetime origin, for example a delta function source $j(y) \propto \delta^{(4)}(y)$, the integral in Eq.~(\ref{eq:theamp}) vanishes.  The ``production vertex" 
for the outgoing wave packet, which is identified spatially with the origin, cannot occur before the time at which the source excites the system.

\begin{figure}[t]
  \begin{center}
    \includegraphics[width=0.5\textwidth]{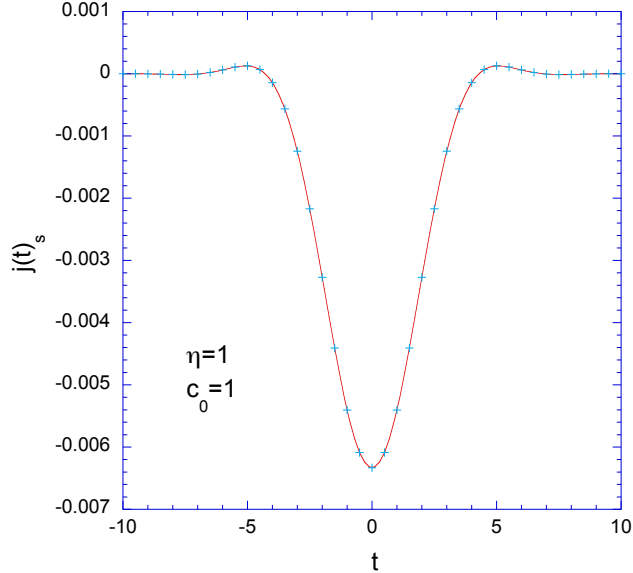}
    \caption{Time-dependence of the smeared source function at the origin $j(t)_s \equiv j(y^0=t,\vec{y}=\vec{0})_s$ for $\eta=1$ and $c_0=1$.}
        \label{fig:js}
  \end{center}
\end{figure}

The conclusion is different if we introduce a nonlocal coupling to the source following our earlier prescription
\begin{equation}
{\cal L }_{int}= [\hat{F}^{1/2} \phi(x)] j(x) \,\,\, .
\end{equation}
This case is simple to understand since we can integrate by parts, and recover a theory of the original form, but with a ``smeared" source,
\begin{equation}
j(y)_s=\hat{F}^{1/2} j(y) = \int d^4 x \, \epsilon(y-x) \, j(x) \,\,\,\,\, \mbox{ where } \,\,\,\,\, \epsilon(y-x) = \int \frac{d^4 k}{(2 \pi)^4} e^{-\eta \, k^4/2} \, e^{i k\cdot(y-x)}  \,\,\,.
\end{equation}
Assuming the example where $j(y)=c_0 \, \delta^{(4)}(y)$, where $c_0$ is a coupling, consider the time-dependence of $j(y)_s$ near the spatial origin
\begin{eqnarray}
j(y^0,\vec{y}=0)_s &=& \frac{c_0}{2 \pi^3} \int_0^\infty dk^0 \int_0^\infty dk\, k^2 \exp[-\eta\, ({k^0}^2-k^2)^2/2] \cos(k^0 y^0) \nonumber \\
&=& \frac{c_0}{8 \sqrt{2} \pi^3} \int_0^\infty dk^0 \, {k^0}^3 e^{-\eta \, k0^4/4} \cos[k^0 y^0]  \left[ K_{1/4}({k^0}^4)-K_{3/4}({k^0}^4) \right]  \,\,\, ,
\label{eq:kint}
\end{eqnarray}
where $K_i$ is a Bessel function of the second kind, of order $i$.  Unlike the original $j(x)$, this function is no longer localized in time at
$t=0$.   The second line of Eq.~(\ref{eq:kint}) can be evaluated numerically and the results are shown in Fig.~\ref{fig:js}.   This result implies that 
there is a common region with $y^0 \ll 0$ and $\vec{y} \approx \vec{0}$ where the functions $j$ and $\tilde{g}$ in Eq.~(\ref{eq:theamp}) both have support; the 
overlap $\langle \psi_{out} | \Omega \rangle$ is therefore generally nonvanishing.   One concludes that there is a non-vanishing probability that
the wave packet appears to emerge from the position of a spatially localized source at a time before the system has
been excited by the source.

{\em Two-into-two scattering.}  The previous example is perhaps the simplest illustration of how the smearing of interaction positions can lead to the apparent
acausal ordering of scattering events.   In the more phenomenologically relevant case of two-into-two scattering, similar results can be obtained. 
We use the term ``apparent" since the spacetime positions of the production and the subsequent decay of a resonance, for example, are inferred from 
the wave packet positions and momenta long before and after the interaction has occurred.  Nothing meaningful can be said about the system directly at intermediate times, 
since no measurements are made during this period.  We now consider how the wave-packet analysis of two-into-two scattering discussed 
in the context of Lee-Wick theories in Ref.~\cite{Grinstein:2008bg} is modified in the present context.

We consider the same free theory of a real scalar field $\phi$, and introduce couplings to two additional real scalar fields $\psi$ and $\chi$, that would otherwise have no
higher-derivative couplings.  In this case we assume
\begin{equation}
{\cal L}_{int} = \frac{1}{2} g_\chi (\hat{F}^{1/2} \phi) \chi^2 + \frac{1}{2} g_\psi (\hat{F}^{1/2} \phi) \psi^2 \,\,\, ,
\label{eq:moddef}
\end{equation}
Again, this is consistent with the assumption that we start with a theory in which the higher-derivative operators appear in the $\phi$ 
quadratic terms only, and that these terms have been subsequently put in canonical form by a field redefinition.  We do not consider
doing the same for the $\psi$ and $\chi$ fields to simplify the analysis; there is no reason to expect that this choice will affect our conclusions
qualitatively. We consider the scattering process $\chi \chi \rightarrow \psi \psi$.   Setting the problem up in the way that we have is convenient since the 
nonlocality affects the propagator but not the external lines, which allows us to immediately carry over most of the wave-packet analysis of 
Ref.~\cite{Grinstein:2008bg} without modification.   Let us briefly recapitulate the key steps in this approach.

We assume incoming and outgoing wavepacket states given by
\begin{align}
| \psi_{in} \rangle & = \int d^4 x_1 d^4 x_2 f_1(x_1-y_1) f_2(x_2-y_2) \phi(x_1) \phi(x_2) | \Omega \rangle  \,\,\ ,\\
| \psi_{out} \rangle & = \int d^4 x'_1 d^4 x'_2 g_1(x'_1-y'_1) g_2(x'_2-y'_2) \phi(x'_1) \phi(x'_2) | \Omega \rangle \,\,\,,
\end{align}
where the functions $f_i$ and $g_i$ define the wavepackets.   These are chosen so that in the process of interest, we can
specify well-defined production and decay vertices for the resonance, in this case associated with the field $\phi$, 
exchanged in the $s$-channel.   To be more explicit, the functions $f_i$ are chosen so that the initial wavepackets are localized
about the space-like separated points $y_1$ and $y_2$, respectively, and have momenta peaked at 
$p_1= m_\chi v_1$ and $p_2 = m_\chi v_2$.  A production vertex can be defined at point $z_0$, where
\begin{equation}
\frac{z_0-y_1}{\tau_1} = v_1 \,\,\,\,\, \mbox{ and } \,\,\,\,\, \frac{z_0-y_2}{\tau_2}=v_2 \,\,\, ,
\end{equation}
with $ \tau_i^2 \equiv (z_0 - y_i)^2$.  Similarly,  the functions $g_i$ are chosen so that the final wavepackets are localized
about the space-like separated points $y'_1$ and $y'_2$, respectively, and have momenta peaked at 
$p'_1= m_\psi v'_1$ and $p'_2 = m_\psi v'_2$.  A decay vertex can be defined at point $z_0'$ where
\begin{equation}
\frac{y'_1-z'_0}{\tau'_1} = v'_1 \,\,\,\,\, \mbox{ and } \,\,\,\,\, \frac{y'_2-z'_0}{\tau'_2}=v'_2 \,\,\, ,
\end{equation}
with ${\tau'}_i^2 \equiv (y'_i-z'_0)^2$.  Defining $w^\mu \equiv {z'}_0^\mu -z_0^\mu$, the authors of Ref.~\cite{Grinstein:2008bg}
determined how the amplitude  $\langle \psi_{out} | \psi_{in} \rangle$ depends on $w^0$ and showed in Lee-Wick
theories that the amplitude is non-vanishing for an acausal ordering of the vertices.  The key intermediate steps are these: the 
amplitude of interest can be written in the form
\begin{equation}
\langle \psi_{out} | \psi_{in} \rangle = \int \frac{d^4 q}{(2 \pi)^4} \tilde{F}(q) \tilde{G}(q) \Gamma^{(4)}_s(q^2) \,\,\, ,
\end{equation}
where $\Gamma^{(4)}_s(q^2)$ is the momentum-space four-point function for the $s$-channel process of interest, with propagators truncated from the external lines, and
\begin{align}
\tilde{F}(q) & = \int d^4 z \, e^{i z \cdot q} \, I_1(z) I_2 (z) \,\,\,\,\, \mbox{ with}\\
I_i(z) & = \int \frac{d^4 k_i}{(2 \pi)^4} \, e^{i k_i \cdot (y_i -z)} \tilde{f}_i(k_i) \, \frac{i}{k_i^2 - m^2 +i \epsilon} \,\,\,,
\label{eq:spints}
\end{align}
where $\tilde{f}_i(k)$ are the Fourier transforms of the incoming wave packet functions.  We do not display the analogous expressions for $G(q)$, corresponding to the outgoing 
wave packet states.  In the same limit described in our earlier example involving a classical source, the momentum and $z$ integrals in Eq.~(\ref{eq:spints}) can be evaluated 
in the stationary phase approximation, leading to a result of the form
\begin{equation}
\langle \psi_{out} | \psi_{in} \rangle \simeq \int \frac{d^4 q}{(2 \pi)^4} \, e^{-i q \cdot (z_0'-z_0)} \, \hat{F}(q)\, \hat{G}(q) \, \Gamma^{(4)}_s(q^2) 
\,\,\, ,
\label{eq:gowres}
\end{equation}
where the functions $\hat{F}$ and $\hat{G}$ have localized support at $q \approx p_1+p_2$ and $q \approx p'_1+p'_2$, respectively.  
We study the nonlocal theory of interest using Eq.~(\ref{eq:gowres}) as the starting point.  Hence, from Eq.~(\ref{eq:moddef})
it follows that
\begin{equation}
\langle \psi_{out} | \psi_{in} \rangle = \int \frac{d^4 q}{(2 \pi)^4} \, e^{-i q \cdot w} 
\left[\frac{- i \,g_\chi \, g_\psi \, e^{-\eta \, q^4} }{q^2-m_\phi^2+i \, m\, \Gamma} \right] \,
\hat{F}(q)\, \hat{G}(q)  \,\,\,,
\label{eq:ourap}
\end{equation}
where $\Gamma$ is the $\phi$ decay width.  Defining the Fourier transform 
\begin{equation}
\epsilon'(x)=\int\frac{d^4 q}{(2\pi)^4} \, e^{-\eta \,q^4} \, e^{i q \cdot x} \,\,\,\, ,
\end{equation}
the amplitude can be written as
\begin{equation}
\langle \psi_{out} | \psi_{in} \rangle = \int d^4 x \, \epsilon(x-w) I(x) \,\,\, ,
\label{eq:epi}
\end{equation}
where
\begin{equation}
I(x) = \frac{\sqrt{i} g_\phi g_\chi}{{8 \pi^{3/2}}} \frac{m_\phi^{1/2}}{(\sqrt{x^2})^{3/2}} e^{-i \, m_\phi \sqrt{x^2}} \, 
e^{-\Gamma \sqrt{x^2}/2} \hat{F}(m_\phi \frac{x}{\sqrt{x^2}}) \, \hat{G}(m_\phi \frac{x}{\sqrt{x^2}})  \,\,\, .
\label{eq:iox}
\end{equation}
As in the case of the ``ordinary resonance" discussed in Ref.~\cite{Grinstein:2008bg}, as well as in our previous example, 
$I(x)$ is derived by exponentiating the propagator denominator using a Schwinger parameter and
then integrating using the stationary phase approximation.  Note that we cannot apply this approximation to Eq.~(\ref{eq:ourap})
directly since we require that the nonlocal length scale $\eta^{1/4}$ to be comparable to the vertex separation; 
one cannot then assume that the real exponential prefactor is a slowly varying function of the momentum relative to the complex phase 
factor.  Eq.~(\ref{eq:iox}) coincides with the ordinary resonance result when $\eta=0$, in which case $\epsilon'$ becomes a four-dimensional 
delta function.  In that limit, the argument of Ref.~\cite{Grinstein:2008bg} is the following: in the center-of-mass frame, $\hat{F}$ and $\hat{G}$ only have support
where $x^0=w^0>0$ and $\vec{x}=\vec{w} \approx0$.  Hence the amplitude is only nonvanishing for the causal
ordering of the production and decay vertices. (In the Lee-Wick case, the result is the opposite.)  Making the same
assumptions here, one concludes only that $x^0$ must be greater than zero;  the amplitude may be nonvanishing, for
example, when $w^0<0$ and $\vec{w}=0$, since $\epsilon'(x-w)$ is no longer a delta function.  This can be verified by noting
that the function $\epsilon'$ differs from the function $\epsilon$ that we have previously studied by the replacement 
$\eta \rightarrow 2 \eta$.  Since the $x$ integral
is dominated by the region where $\vec{x}\approx 0$, we can evaluate $\epsilon(x-w)$ for the choice $\vec{x}=\vec{w}=0$,
where $\vec{w}=0$ corresponds to the case in which the production and decay vertices are spatially coincident. Since $\epsilon'$
is non-vanishing for $x^0>0$ and $w^0<0$ we conclude that $\epsilon'$, $\hat{F}$ and $\hat{G}$ have common regions of support,
so that Eq.~(\ref{eq:epi}) is generally non-vanishing.  Hence, there is a non-vanishing amplitude for an acausal ordering of the production 
and decay vertices.  The effect emerges in a very different way than in the Lee-Wick theories.  In that case, a crucial sign flip in the propagator of the
Lee-Wick resonance leads to a change from $w^0$ to $-w^0$ in comparison to the ordinary resonance case.  The sign flip
affects the sign of the width appearing in one of the exponential factors in the amplitude, leading to the interpretation that
the exponential decay is happening as the Lee-Wick resonance propagates backward in time from the decay to production
vertex.  Here, however, the form of $I(x)$ corresponds to propagation forward in time over the time-like interval $x$.  The 
nonlocality in the theory leads to a spatial smearing of the interaction points so that one no longer identifies $x$ with the physical 
spacetime separation of the extrapolated decay and production vertices.

\section{Conclusions} \label{sec:conc}

We have considered unitarity and causality in a theory where quadratic terms are modified by higher-derivative terms of
infinite order, chosen so as not to induce additional poles in the propagator at tree level.   We have studied unitarity at leading order in the large $N$ limit of the
scalar O($N$) model for Euclidean and Minkowski space formulations of the theory.   We have verified that a unitary theory
is obtained from Euclidean correlation functions that are analytically continued in their external momenta to Minkowski space,
but not when correlation functions are formulated in Minkowski space directly.  In the same theory, we verified the non-vanishing of field 
commutators at space-like separation by an explicit calculation.  We then studied the time-dependence of scattering amplitudes in similar theories
using a wave-packet  approach employed by others~\cite{Grinstein:2008bg} in studying Lee-Wick theories.  We found that the apparent acausal ordering of 
decay and production vertices of resonances was a common feature in these theories.  Unlike Lee-Wick theories, this effect would be present in tree-level resonant 
exchanges for all the states in the theory that are subject to modified quadratic terms and would make solutions to the hierarchy problem based on this idea phenomenologically distinct from others 
that have been proposed.   
%%%%%%%%%%%%%%%%%%%%%%%%%%%%%%%%%%%%%%%%%%%%%%%%%%%%%%%%%%%
\begin{acknowledgments}  
We thank Josh Erlich for valuable discussions. This work was supported by the NSF under Grant PHY-1519644, and is dedicated to the memory of 
David A. Carone.
\end{acknowledgments}
%%%%%%%%%%%%%%%%%%%%%%%%%%%%%%%%%%%%%%%%%%%%%%%%%%%%%%%%%%%     

\appendix
\section{Schwartz Reflection} \label{sec:appendix}

In the text, we computed the imaginary part of $\Sigma(p^0)$ directly.  If one instead were to compute the discontinuity
about the cut along the real axis using the usual Cutkosky cutting rules, one would obtain a unitary theory only if the relation between
the discontinuity and the imaginary part were determined by $\Sigma(p^0)=\Sigma(p^{0*})^*$.   This property is called the 
Schwartz reflection principle.  Our numerical results in Sec.~\ref{sec:unitarity}B suggest that there is no interval along the real $p^0$
axis where this relation is valid.  In this appendix, we show that the condition that $\Sigma(p^0)=\Sigma(p^{0*})^*$ is identical to the requirement 
that the relevant $k^0$ loop integral about a contour at large radius in the complex plane vanishes identically, which is not the case in the 
theory defined in Minkowski space.

\begin{figure}[t]
 \begin{center}
   \includegraphics[width=0.5\textwidth]{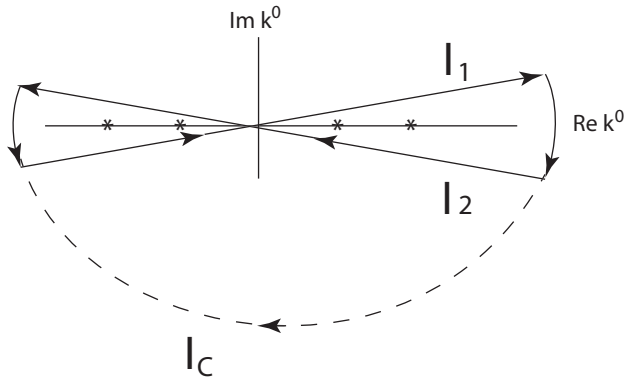}
\caption{Contours discussed in the Appendix for studying the reality properties of $\Sigma(p^0)$ for real $p^0 < 2 \, m$ and $\vec{p}=0$. The contour
 for $I_{C}$ terminates on the $I_1$ line.  \label{fig:afig}}      
  \end{center}
\end{figure}

To illustrate this, consider real $p^0 < 2 \, m$ with $\vec{p}=0$.  Let
\begin{equation}
I(p^0) = -\frac{i}{2} \int\frac{d^3 k}{(2\pi)^3} \int \frac{dk^0}{(2 \pi)} \frac{B(k^0,\vec{k})}{\left[(k^0 - p^0/2)^2-E_{\vec{k}}^2\right]\left[(k^0 + p^0/2)^2-E_{\vec{k}}^2\right]}\,\,\, ,
\label{eq:variousi}
\end{equation}
where $B(k^0,\vec{k})$ represents the numerator factor in Eq.~(\ref{eq:sigmink}).   For real $p^0 < 2 \, m$, the usual Feynman prescription calls for  
going below the poles at $k^0 = \pm p^0/2 - E_{\vec{k}}$ (both in the left half-plane) and above those at $k^0 = \pm p^0/2 + E_{\vec{k}}$ (both in
the right half-plane).   We achieve this by evaluating the integral along a contour defined by $k^0 = \rho\, e^{i \, \epsilon}$, for a real integration variable $\rho$, 
and then taking the limit as $\epsilon$ approaches zero.  Hence, the integral labelled $I_1$ in Fig.~\ref{fig:afig} is identical to the function 
$\Sigma(p^0)$ discussed earlier.  On the other hand, $\Sigma(p^{0*})^*$ (again assuming real $p^0$) corresponds to the same integral evaluated along 
the path $k^0 = \rho\, e^{-i \, \epsilon}$, but in the opposite direction due an additional overall minus sign from complex conjugation. This is the integral $I_2$ shown 
in the figure.     

Integration along either of the the two arcs at large radius shown in Fig.~\ref{fig:afig} is identically zero, since the function $B$ is damped as one 
approaches the real axis, even when the nonlocality parameter $\eta$ is nonzero.   Let us denote the residues of the four poles shown in 
Fig.~\ref{fig:afig} by $R_i$ for $i=1\ldots 4$ going from left to right.   Considering the two closed, wedge-shaped loops shown in the figure, it follows 
from the residue theorem that 
\begin{equation}
I_1 + I_2  = 2 \pi i \, (R_1+R_2 -R_3 - R_4)  \,\,\, .
\end{equation}
However, using the fact that the function $B$ is even in $k^0$, one may compute the residues directly and confirm that $R_1 = -R_4$ and $R_2 = -R_3$.
Hence,  
\begin{equation}
I_1 + I_2  = -4 \pi i \, (R_3+R_4) \,\,\, .
\label{eq:almost}
\end{equation}
Next, consider the semi-circular contour in the lower-half plane that terminates on the $I_1$ contour.  Clearly, $I_1+I_{C} =-2 \pi i \, (R_3+R_4)$.  Combining
this with Eq.~(\ref{eq:almost}) it follows that $I_1 + I_2 = 2 (I_1 + I_{C})$, or using our previous identification:
\begin{equation}
\Sigma(p^0) + 2 \, I_{C}(p^0) = \Sigma(p^{0*})^* \,\,\,.
\end{equation}
In the case where $B=1$, the integrand of Eq.~(\ref{eq:variousi}) drops off in all directions in the complex $k^0$ plane.  Hence, $I_{C}=0$, and the
relation $\Sigma(p^0)=\Sigma(p^{0*})^*$ is obtained; it can then be analytically continued to complex $p^0$ to relate the discontinuity to the 
imaginary part. In the theory studied in Sec.~\ref{sec:unitarity}B, there is no general reason to expect that $I_C$ is nonvanishing (the integrand
grows in certain directions in the complex plane) and it is the same as the integral $I_C$ discussed in that section that was found to be non-zero by direct 
numerical evaluation. In this case, it is not justified to analytically continue  $\Sigma(p^0)=\Sigma(p^{0*})^*$ to determine the relation between 
the discontinuity and the imaginary part.

\end{document}